\documentclass[aps,twocolumn,pra,showpacs,floatfix]{revtex4}
\usepackage{dcolumn}

\begin{document}

\title{\bf Scalar Static Polarizabilities of Lanthanides and
  Actinides.} 
\author{V. A. Dzuba, A. Kozlov,  and V. V. Flambaum}
\affiliation{School of Physics, University of New South Wales, 
Sydney 2052, Australia}
\date{\today}

\begin{abstract}
We calculate scalar static polarizabilities for lanthanides and
actinides, the atoms with open $4f$ or $5f$ subshell. We show that
polarizabilities of the low states are approximately the same for all
states of given 
configuration and present a way of calculating them reducing valence
space to just two or three valence electrons occupying $6s$ and $5d$
states for lanthanides or $7s$ and $6d$ states for actinides while
$4f$ and $5f$ states are considered to be in the core. Configuration
interaction technique is used to calculate polarizabilities of
lanthanides and actinides for all states of the $4f^n6s^2$ and
$4f^{n-1}6s^25d$ configurations of lanthanides and all states of the
$5f^{n}7s^2$ and $5f^{n-1}7s^26d$ configurations of actinides.
Polarizability of the electron core (including f-orbitals) has been
calculated in the RPA approximation. 
\end{abstract} 
\maketitle

\section{introduction}

The main characteristic of a neutral atom which determines its
interaction with the environment is its polarizability. The van der
Waals forces between atoms, atom-wall interaction, interaction of
neutral atoms with laser electric field in an optical lattice are all
related to polarizabilities (see, e.g.  \cite{Bloch:2005,Bloom:2014}).  
Interest to accurate measurement and calculations of atomic
polarizabilities rose over the last decade with  the development of
next generation of atomic clocks based on optical transitions
\cite{Rosenband:2007, Yb}. Accuracy of optical clocks is mostly
limited by the blackbody radiation shift (BBR) (see,
e.g. \cite{Rosenband:2006,   
Madej:2012, Porsev:2006, Heavner:2005})  which is
proportional to differential polarizability of two atomic clock
states. There is a review by Mitroy {\em et
  al}~\cite{Mitroy:2010} which describes in detail current status of
the experimental and theoretical study of atomic polarizabilities. In
brief, it is as follows. Polarizabilities are well studied for ground
states of noble gases and for ground and some excited states of
atoms with simple electron structure, i.e. atoms which have one, two
or three valence electrons above closed shells. Experimental data for
excited states is poor. This is one of the motivations for accurate
atomic calculations.  Having accurate values of atomic
polarizabilities for excited states is important for estimation of the
BBR shift and for finding the so called {\em magic} frequencies of
laser field that makes optical lattice for which the electric dynamic
energy shift of both clock states are the same so that clock frequency
is not affected by lattice field.

In contrast to relatively rich data for atoms with simple electron
structure, the situation for atoms with open $d$ or $f$ shells is very
much different. Apart from very few exceptions, the experimental data
is practically absent. Theoretical data is presented by a single
unpublished work by Doolen~\cite{Doolen} which, in spite of being
unpublished, is widely cited in textbooks and
databases ( see, e.g. \cite{Miller,Peter}). It uses a relativistic
linear response method~\cite{Zangwill}, estimated uncertainty is 25\%. 

Knowing polarizabilities of open-shell atoms is important for many
applications. For example, it was suggested in Ref.~\cite{DFG2010} to
search for positron-atom bound states through resonant annihilation.
The method would work for atoms with open shells which have low-lying
excited states within the ground state
configuration~\cite{DFGH2012,HDF2014}. Kinetic energy of scattering
positron is spent on exciting the atom and positron is bound to the
exciting state. Polarizability is an important characteristic of the
atoms governing their ability to bind a positron. In this paper we
argue that polarizabilities of all states of the same configuration are
approximately the same. Therefore, if positron is bound to the ground
state it is very likely to be bound to an excited state of the same
configuration. 

Lanthanides and actinides are also used in many other important
studies. For example, Yb and Er are considered for very precise atomic
clocks~\cite{Yb,Er}; parity non-conservation has been measured in
Dy~\cite{Leefer:2013} and Yb~\cite{Tsigutkin}; Dy and Er are used to
study quantum gases~\cite{Dy, Innsbruck}; Th is considered for
ultra-precise nuclear clock~\cite{Th}, etc. The heaviest of the
actinides approach an important area of superheavy
elements~\cite{Pershina}. In terms of electron structure, there is
practically no experimental data for superheavy elements, all data
comes from theory and polarizability is one of the most important
characteristics. 

In this paper we try to address the lack of data on polarizabilities
of lanthanide and actinides. We propose a method of calculation which
reduces the calculation of polarizabilities of lanthanides and
actinides to calculations for a system with two or three valence
electrons. The approach is based on the assumption that residual
Coulomb interaction between $f$ and other valence electrons is small
so that total angular momenta of each subsystem are still good quantum
numbers. This allows us to attribute $f$-electrons to the core
reducing the problem to calculation of the polarizabilities of the
$6s^2$ or $6s^25d$ configurations of the valence electrons for
lanthanides and $7s^2$ or $7s^26d$ configurations for actinides. To
check the approach we have perfumed test calculations for few systems
in which $f$-electrons were treated as valence states. The agreement
between two approaches is very good. There is also surprisedly  
good agreements with early calculations by Doolen~\cite{Miller}. As a
rule, the difference between our results and those of
Doolen~\cite{Miller} is much less that the 25\% uncertainty claimed in
\cite{Miller}. There is also good agreement with the experimental data
for uranium. However, we have significant disagreement with the
results of the measurements of the dynamic polarizabilities for
Dy~\cite{Lev} and Er~\cite{Ferlaino}. The possible reasons for this
disagreement are discussed. 

\section{General formalism}

Second-order Stark shift of atomic energy level in static electric
field $\mathcal{E}$ can be written as
\begin{equation}
  \Delta E_a = -\frac{1}{2}\alpha(a) \mathcal{E}^2,
\label{eq:Stark}
\end{equation}
where polarizability $\alpha$ is the sum of scalar and tensor terms
\begin{equation}
  \alpha(a) = \alpha_0(a) +\frac{3M_a^2-J_a(J_a+1)}{J_a(2J_a-1)}\alpha_2(a).
\label{eq:Stark}
\end{equation}
Here $J_a$ is the total angular momentum of the atom and $M_a$ is its
projection on the direction of the electric field.
Scalar polarizability $\alpha_0(a)$ and tensor polarizability $\alpha_2(a)$
can be expressed via sums over complete sets of intermediate states
involving matrix elements of the electric dipole operator $\mathbf{D}$
(in length form $\mathbf{D} = -e\sum_i \mathbf{r}_i$)
\begin{eqnarray}
&&\alpha_0(a) = \frac{2}{3(2J_a+1)} \sum_n \frac{\langle a||\mathbf{D}|| n
  \rangle^2}{E_a - E_n},
\label{alpha0} \\
&&\alpha_2(a) = 2\sqrt{\frac{10J_a(2J_a-1)}{3(2J_a+3)(2J_a+1)(J_a+1)}}
\times \nonumber \\
&&\sum_n (-1)^{J_a+J_n} \left\{ \begin{array}{lll} 1 & 1 & 2 \\ J_a &
    J_a& J_n \end{array} \right\} \frac{\langle a||\mathbf{D}|| n 
  \rangle^2}{E_a - E_n}.
\label{alphaq}
\end{eqnarray}
Here $|a\rangle$ and $|n\rangle$ are many-electron atomic states and
$E_a$ and $E_n$ are corresponding energies. Tensor polarizability
(\ref{alphaq}) is none-zero for $J_a \geq 1$ while scalar
polarizability is none-zero even for $J_a=0$. 

In this paper we consider only scalar polarizabilities.

\subsection{Polarizabilities of closed-shell atoms}

\label{s:c-s}

\begin{table}\center
\caption{Comparison of calculations of scalar polarizabilities of some
  noble gases with experimental values presented in
  \cite{Mitroy:2010}. Values are in atomic units.} 
\label{t:noble}
\begin{tabular}{c | c | c}
\hline\hline
&&\rule{0pt}{4pt}\\
element & calculation & experiment\\
&&\rule{0pt}{2pt}\\
\hline
&&\rule{0pt}{2pt}\\
Ar & 10.77 & 11.08\\
&&\rule{0pt}{2pt}\\
\hline
&&\rule{0pt}{2pt}\\
Kr & 16.47 & 16.74\\
&&\rule{0pt}{2pt}\\
\hline
&&\rule{0pt}{2pt}\\
Xe & 26.97 & 27.34\\
&&\rule{0pt}{4pt}\\
\hline\hline
\end{tabular}\label{Tab:5}
\end{table}

\begin{table}\center
\caption{Contributions to scalar polarizabilities of some
  atoms with open $f$-shell from core states (below the $4f$ or $5f$
  states), $4f$ ($5f$), and $6s$ ($7s$) states. Values are in atomic units.} 
\begin{tabular}{c | c | c | c | c}
\hline\hline
&&&&\rule{0pt}{4pt}\\
element & core & $4f^{N-2}(5f^{N-2})$ & $6s^2(7s^2)$ & Total\\
&&&&\rule{0pt}{2pt}\\
\hline
&&&&\rule{0pt}{2pt}\\
Dy & -3.3& -1.9& 215& 209.8\\
&&&&\rule{0pt}{2pt}\\
\hline
&&&&\rule{0pt}{2pt}\\
Er & -3& -2.1& 195.4& 193.3\\
&&&&\rule{0pt}{2pt}\\
\hline
&&&&\rule{0pt}{2pt}\\
Yb & -2.6& -2.5& 183.7& 178.6\\
&&&&\rule{0pt}{2pt}\\
\hline
&&&&\rule{0pt}{2pt}\\
Pu & -2 & -2 & 216.6 & 212.6\\
&&&&\rule{0pt}{4pt}\\
\hline\hline
\end{tabular}\label{Tab:6}
\end{table}

For closed-shell atoms tensor polarizability is zero and scalar 
polarizability is given by
\begin{equation}
\alpha_0 = \frac{2}{3} \sum_n \frac{\langle a||\mathbf{D}|| n
  \rangle^2}{E_0 - E_n}.
\label{alpha0cs}
\end{equation}

In the random-phase approximation (RPA) expression (\ref{alpha0cs}) 
is reduced to the sum over single-electron matrix elements
\begin{equation}
\alpha_0 = \frac{2}{3} \sum_{cn} \frac{\langle c||\mathbf{d}+\delta
  \mathbf{V}|| n 
  \rangle\langle n||\mathbf{d}|| c \rangle}{\epsilon_n - \epsilon_c},
\label{alpha0RPA}
\end{equation}
where $\mathbf{d} =-e\mathbf{r}$ is the single-electron electric dipole
operator, $\delta\mathbf{V}$ is correction to the core potential due to
core polarization by external electric field; summation goes over core
states $c$ and complete set 
of single-electron orbitals $n$. The energies $\epsilon_c$ and
$\epsilon_n$ are the Hartree-Fock energies of single-electron
orbitals $n$ and $c$. Note that the core polarization correction
$\delta \mathbf{V}$ is included in one of the electric dipole matrix 
elements only. This is because for a closed-shell system there is
only one infinite chain of RPA diagrams standing between two electric
dipole operators. It can be attributed
to one of the operators but not to both \cite{Pendrill}.

The RPA approximation (\ref{alpha0RPA}) gives good accuracy
for noble gases (see Table \ref{t:noble}). It is also
sufficiently accurate for the polarizabilities of closed-shell
atomic cores. It is widely used in the calculations of atomic
polarizabilities in which core and valence contributions are 
calculated separately and then added together.

Formally, Eq.~(\ref{alpha0RPA}) can  
be used for any closed-shell systems, such as e.g. Ba, Yb, etc.
It can be even used for open-shell systems if fractional
occupation numbers formalism is used. However, the calculated
RPA polarizability of such systems is usually overestimated. 
This this due to neglecting of important contribution of
inter-electron correlations. Correlations produce additional
attraction between electrons making the atom to be more compact
and reducing its polarizability. The RPA calculations can still 
be used for rough estimations and for the study of relative 
contributions of different atomic subshells. Table \ref{Tab:6},
in which RPA polarizabilities of f-elements are presented, shows
that the polarizabilities of f-elements are strongly dominated 
by external 6s- and 5d-electrons while the contribution of 
4f-electrons is small. This means that the correlations should
be treated accurately for two or three valence electrons while
they can be neglected in other contribution. Inclusion of correlations
is discussed in section \ref{Method}.

Note that the contribution of the $f$-states to the polarizability is
negative (as well as the total contribution of the lower core
states). It may look as an unexpected result since all terms in the
exact expression (\ref{alpha0cs}) are positive. Total polarizability
of the ground state is always positive. This is just a reflection of
the well known fact that the second-order perturbation correction to
the energy, which is related to polarizability via (\ref{eq:Stark}),
is always negative. However, in the RPA approximation
(\ref{alpha0RPA}) only total polarizability is positive. Partial
contributions might be negative due to the different sign of the
$\langle c||\mathbf{d}+\delta \mathbf{V}|| n \rangle$ and
$\langle c||\mathbf{d}|| n \rangle$ matrix elements. This only happens
for lower states in the core and can be explained by screening of the
external electric field in atoms~\cite{DFSS86}. The screened field has
complex oscillating behavior inside atomic core often having different sign on
wide range of distances. Note that screening is treated pretty
accurately in the RPA approximation, e.g. Schiff theorem (complete
screening of external electric field by electrons at the nucleus of an
atom) fulfills exactly~\cite{DFSS86}. 

\subsection{Polarizabilities of compound systems}

\label{compound}

To derive a way of calculating polarizabilities of complicated
many-electron systems we start from a very general statement. If the
system can be divided into two subsystems so that the total wave
function is the product of wave functions of each subsystem connected
by Clebsh-Gourdan coefficient then the polarizability of the whole
system is the sum of polarizabilities of two subsystems. 
Such presentation is possible when residual Coulomb interaction
between electrons of the two subsystems is small.

A case when the total angular momentum of one of the subsystems is
zero is widely used in the calculations of the atomic
polarizabilities. The total polarizability is presented as a sum of
the contributions from closed-shell atomic core and from valence
electrons. These contributions are calculated separately and then
added together. Note that there are also cross contributions caused by
Pauli principle. Calculation of polarizabilities of one subsystem
is affected by the other subsystem. States occupied by electrons of
other system must be excluded from the summation over intermediate
states due to Pauli principle. These contributions are usually small
and we will ignore them in our consideration. There are also
cancellations between Pauli-forbidden contributions to each of the
polarizabilities. 

We will consider a non-trivial case when total angular momentum of
both subsystems is not zero. The wave function of the whole system is 
\begin{equation}
|a \rangle = \sum_{M_1,M_2} C^{J_aM_a}_{J_1M_1J_2M_2}|a'
J_1M_1\rangle |a''J_2M_2 \rangle, \label{eq:a} 
\end{equation}
where $J_a, M_a$ are the total angular momentum of the system and its
projection,  $J_1, M_1$ and $J_2, M_2$ are total angular momenta and
projections for each subsystem, $C^{JM}_{J_1M_1J_2M_2}$ is the
Clebsh-Gourdan coefficient.   

The electric dipole operator $\mathbf{D}$ in the expression
(\ref{alpha0}) for the scalar polarizability can be
written as a sum $\mathbf{D}=\mathbf{D}_1+\mathbf{D}_2$ in which
summation in $\mathbf{D}_1$ goes over 
electrons of first subsystem and summation in $\mathbf{D}_2$ goes over
electrons of second subsystem. Let us consider the contribution of
$\mathbf{D}_2$ to the polarizability (\ref{alpha0}). States $|n\rangle$ which
contribute to the polarizability can be written as
\begin{equation}
|n \rangle = \sum_{M_1,M_3} C^{J_nM_n}_{J_1M_1J_3M_3}|n'
J_1M_1\rangle |n''J_3M_3 \rangle \label{eq:n}. 
\end{equation}
Here first part of the wave function is the same as in (\ref{eq:a})
and second part satisfies selection rules for electric dipole
transition between states $|a''\rangle$ and $|n''\rangle$, they have
opposite parity and $J_3=J_2,J_2 \pm 1$.

Substituting (\ref{eq:a}) and (\ref{eq:n}) into the square of the
electric dipole matrix element we get 
\begin{eqnarray}
&&\langle a ||\mathbf{D}|| n\rangle^2 = \left( \begin{array}{rrr} J_a
    & 1 & J_n \\ 
    -M_a & 0 & M_n \end{array} \right)^{-2}  \times \nonumber \\
&&\left[ \sum_{M_1,M_2,M_3}  C^{J_aM_a}_{J_1M_1J_2M_2}
  C^{J_nM_n}_{J_1M_1J_3M_3} (-1)^{J_2-M_2} \times \right.\nonumber \\ 
&& \left. 
  \left( \begin{array}{rrr} J_2 & 1 & J_3 \\ -M_2 & 0 & M_3 \end{array}
\right) \right]^2 
\langle a'' J_2 ||\mathbf{D}|| n''  J_3 \rangle^2
= \label{eq:summ} \\ 
&& (2J_a+1)(2J_n+1)\left\{ \begin{array}{lll} J_a & 1 & J_n \\ J_3 & J_1 &
      J_2 \end{array} \right\}^2 
\langle a'' J_2 ||\mathbf{D}|| n''  J_3 \rangle^2. 
\nonumber
\end{eqnarray}
Here formula (12.1.6) from Ref.~\cite{Varshalovich} was used. Noting
that calculation of the polarizability involves summation over
different values of total angular momentum $J_n$ and using
\begin{equation}
\sum_{J_n} (2J_n+1)\left\{ \begin{array}{lll} J_a & 1 & J_n \\ J_3 &
      J_1 & J_2 \end{array} \right\}^2  = \frac{1}{(2J_2+1)}
\label{eq:sumj}
\end{equation}
(see (12.2.15) from Ref.~\cite{Varshalovich}), the expression
(\ref{alpha0}) is reduced to
\begin{equation}
\alpha_0(a'') = \frac{2}{3(2J_2+1)} \sum_{n''} \frac{\langle a''
  J_2||\mathbf{D}||   n'' J_3 \rangle^2}{E_{a''} - E_{n''}}.
\label{alpha2}
\end{equation}
We see that the contribution of $\mathbf{D}_2$ into total
polarizability of the 
system is reduced to calculation of the polarizability of second
subsystem as if there is no first subsystem. Expression (\ref{alpha2})
does not depend neither on the total angular momentum $J_1$ of first
subsystem nor on the total angular momentum $J_a$ of the whole
system.

\subsection{Application to f-elements}

\label{f-elems}

To calculate polarizabilities of f-elements using approach considered 
in previous section we divide all valence electrons into two subsystems,
one has f-electrons only and other has all remaining electrons, namely 
two s-electrons or two s-electrons and one d-electron. We will consider
lanthanides as an example. However, the same consideration is valid 
for actinides as well. 

The wave function lowest states of lanthanides can be written as either
\begin{equation}
|a \rangle = \sum_{M_1,M_2} C^{J_aM_a}_{J_1M_1J_2M_2}|4f^n
J_1M_1\rangle |6s^2 J_2M_2 \rangle, \label{eq:6s2} 
\end{equation}
or
\begin{equation}
|a \rangle = \sum_{M_1,M_2} C^{J_aM_a}_{J_1M_1J_2M_2}|4f^{n-1}
J_1M_1\rangle |6s^25d J_2M_2 \rangle, \label{eq:6s25d} 
\end{equation}
where $J_a$ is the total angular momentum of the atom, $M_a$ 
is its projection,  $J_1, M_1$ are the total angular momentum and 
its projection of the $4f^n$ or $4f^{n-1}$ subsystem,
$J_2, M_2$ are the total angular momentum and its
projection for the $6s^2$ or $6s^25d$ subsystem, 
$C^{JM}_{J_1M_1J_2M_2}$ is the Clebsh-Gourdan coefficient.   
The quality of the approximation (\ref{eq:6s2}) or (\ref{eq:6s25d})
for lanthanides can be illustrated by similarities in the spectra
of neutral atoms and their double (or triple) ionized ions.

Applying the consideration of previous section we see that the 
calculation of polarizabilities of lanthanides is reduced to
calculation of the polarizability of unfilled f-subshell and
the polarizability of the remaining $6s^2$ or $6s^25d$ valence
electrons.

As we have seen in section \ref{s:c-s} the contribution of f-electrons
into polarizability is small. It can therefore be calculated in a
single-configuration approximation
with the use of fractional occupation numbers as discussed in section
\ref{s:c-s}. It is the best to attribute the $4f$  
electrons to the core so that their contribution to the 
self-consistent Hartree-Fock potential and to polarizability
is calculated in a similar way with the use of fractional
occupation numbers.

The dominant contribution to the polarizabilities comes from
valence $6s$ and $5d$ electrons. Its calculation is now 
reduced to the calculation of the polarizability of two
or three valence electrons system. The calculations for the
$4f^n6s^2$ configuration are reduced to the calculations for
the $6s^2$ configuration as for ytterbium \cite{Dzuba:2010}; the
$4f^{n-1}6s^25d$ configuration is reduced to the $6s^25d$ one
as in lutetium. No further approximation is
needed and full power of the configuration interaction technique
combined with the many-body perturbation theory (the CI+MBPT
method~\cite{Dzuba:1996}) can be used. 
The details of the calculations for few valence electron systems can
be found in our earlier works \cite{Dzuba:1996,Dzuba:2005, Ginges:2006}. 

Note that since expression (\ref{alpha2}) does not depend on the
total angular momentum of the atom, the scalar polarizability of the
atom in this approximation is the same for all states of the same
configuration. We have demonstrated this already for erbium in our
previous work~\cite{Er}.

\subsection{Application to d-elements}

One may argue that the approach developed above should also work for
atoms with open $d$-shells. Indeed, some of the supporting arguments 
do work for such atoms. For example, the contribution of the
$d$-states into polarizabilities of atoms with open $d$-shells is
small. However, the more important condition, small value of the
residual Coulomb interaction (see section \ref{compound}), is not
always fulfilled for such atoms. This manifests itself in
configuration mixing and can be verified by examining the spectra of
the open-shell atoms. The states of the $4f^n6s^2$ configuration are
sufficiently pure. Mixing with configurations having different number
of $4f$-electrons is small. This is
because the $4f$ electrons are most easily excited into the $5d$
state, but configurations $4f^n6s^2$ and $4f^{n-1}5d6s^2$ do not mix
due to different parity. On the other hand, the states of the same
parity and total angular momentum but different configurations are
high in the spectrum. For example, the first state of erbium which
could mix with the $4f^{12}6s^2 \ ^3$H$_6$ ground state is the state
of the $4f^{11}6s^26p$ configuration with the energy of
19817~cm$^{-1}$.  

In contrast, the states of the $5d^n6s^2$ configurations are not pure
due to mixing with the $5d^{n+1}6s$ configuration. The same is true
for most of the atoms with the $4d^n5s^2$ or $3d^n4s^2$ ground state
configuration. For example,
the $4d^55s^2 \ ^6$S$_{5/2}$ ground state of technetium is mixed with
the $4d^65s \ ^6$D$_{5/2}$ excited state separated by 3701~cm$^{-1}$
only. There are atoms in which such energy interval is large. The
approach used in this paper might work for these atoms. This question
needs additional study.

\begin{table*}\center
\caption{Scalar polarizabilities of lanthanides. All values are given in atomic units.}
{\renewcommand{\arraystretch}{0}%
\begin{tabular}{l c l l l l l l}
\hline\hline
\rule{0pt}{4pt}\\
Z & element & configuration & core & \vtop{\hbox{\strut
    valence}\hbox{\strut (CI+MBPT)}} & \vtop{\hbox{\strut total
    scalar}\hbox{\strut polarizability}\hbox{\strut (core+valence)}} &
\vtop{\hbox{\strut existing}\hbox{\strut data}} & reference \\ 
\rule{0pt}{4pt}\\
\hline
\rule{0pt}{4pt}\\
57 & La & $5d6s^2$ & 7.7& 206& 213.7 & 210 & calc. \cite{Miller}\\
 & & $5d^26s^1$& 7.7& 211& 218.7 & $-$ & $-$\\
\rule{0pt}{2pt}\\
\hline
\rule{0pt}{2pt}\\
58 & Ce & $4f^15d6s^2$ & 5.5& 199.2& 204.7& 200 & calc. \cite{Miller}\\
 & & $4f^26s^2$& 4.1& 219.3& 223.4& $-$ & $-$ \\
\rule{0pt}{2pt}\\
\hline
\rule{0pt}{2pt}\\
59 & Pr & $4f^36s^2$ & 4.7& 211.1& 215.8 & 190 & calc. \cite{Miller}\\
 & & $4f^25d6s^2$& 5.3& 190.4& 195.7& $-$ & $-$\\
\rule{0pt}{2pt}\\
\hline
\rule{0pt}{2pt}\\
60 & Nd & $4f^46s^2$ &  5.3 & 203.1 & 208.4 & 212 & calc. \cite{Miller}\\
 & & $4f^35d6s^2$& 5.1& 182.4& 187.5 & $-$ & $-$\\
\rule{0pt}{2pt}\\
\hline
\rule{0pt}{2pt}\\
61 & Pm & $4f^56s^2$ & 5.6 & 194.6 & 200.2 & 203 & calc. \cite{Miller}\\
 & & $4f^45d6s^2$& 5 & 174.3& 179.3& $-$ & $-$\\
\rule{0pt}{2pt}\\
\hline
\rule{0pt}{2pt}\\
62 & Sm & $4f^66s^2$ & 5.8 & 186.3 & 192.1 & 194 & calc. \cite{Miller}\\
 & & $4f^55d6s^1$& 4.9& 166.8& 171.7 & $-$ & $-$\\
\rule{0pt}{2pt}\\
\hline
\rule{0pt}{2pt}\\
63 & Eu & $4f^76s^2$& 5.9 & 178.3 & 184.2 & 187 & calc. \cite{Miller}\\
 & & $4f^65d6s^1$& 4.8 & 159.9& 164.7 & $-$ & $-$\\
\rule{0pt}{2pt}\\
\hline
\rule{0pt}{2pt}\\
64 & Gd & $4f^75d6s^2$ & 4.7& 153.6& 158.3 & 159 & calc. \cite{Miller}\\
 & & $4f^75d^26s^1$& 4.7& 189.8& 194.5 & $-$ & $-$\\
\rule{0pt}{2pt}\\
\hline
\rule{0pt}{2pt}\\
65 & Tb & $4f^96s^2$& 6.1 & 163.4 & 169.5 & 172 & calc. \cite{Miller} \\
 & & $4f^85d6s^2$& 4.6& 147.8& 152.4& $-$ & $-$\\
\rule{0pt}{2pt}\\
\hline
\rule{0pt}{2pt}\\
66 & Dy & $4f^{10}6s^2$ & 6.1 & 156.6 & 162.7 & 165 & calc. \cite{Miller}\\
 & & $4f^95d6s^2$& 4.5& 142.8& 148.3& $-$ & $-$\\
\rule{0pt}{2pt}\\
\hline
\rule{0pt}{2pt}\\
67 & Ho & $4f^{11}6s^2$& 6.2 & 150.1 & 156.3 & 159 & calc. \cite{Miller}\\
 & & $4f^{10}5d6s^2$& 4.4& 138.5& 142.9& $-$ & $-$\\
\rule{0pt}{2pt}\\
\hline
\rule{0pt}{2pt}\\
68 & Er & $4f^{12}6s^2$& 6.3 & 143.9 & 150.2 & 153 & calc. \cite{Miller}\\
 & & $4f^{11}5d6s^2$& 4.4& 135& 139.4& $-$ & $-$\\
\rule{0pt}{2pt}\\
\hline
\rule{0pt}{2pt}\\
69 & Tm &$4f^{13}6s^2$ & 6.3 & 138 & 144.3 & 147 & calc. \cite{Miller}\\
 & & $4f^{12}5d6s^2$& 4.3& 132.5& 137.8& $-$ & $-$\\
\rule{0pt}{2pt}\\
\hline
\rule{0pt}{2pt}\\
70 & Yb & $4f^{14}6s^2$& 6.4 & 132.5 & 138.9 & 142 & calc. \cite{Miller}\\
 & & $4f^{14}6s^16p^1$& 6.4 & 305.8 & 312.2 & 315.9 & calc. \cite{Dzuba:2010}\\
\rule{0pt}{2pt}\\
\hline
\rule{0pt}{2pt}\\
71 & Lu & $4f^{14}5d6s^2$& 4.3 & 132.9& 137.2 & 148 & calc. \cite{Miller}\\
 & & $4f^{14}6s^26p^1$& 4.3& 57& 61.3& $-$ & $-$ \\
\rule{0pt}{4pt}\\
\hline\hline
\end{tabular}}\label{Tab:1}
\end{table*}

\begin{table*}\center
\caption{Scalar polarizabilities of actinides. All values are given in atomic units.}
{\renewcommand{\arraystretch}{0}%
\begin{tabular}{l c l l l l l l}
\hline\hline
\rule{0pt}{4pt}\\
Z & element & configuration & core & \vtop{\hbox{\strut
    valence}\hbox{\strut (CI+MBPT)}} & \vtop{\hbox{\strut total
    scalar}\hbox{\strut polarizability}\hbox{\strut (core+valence)}} &
\vtop{\hbox{\strut existing}\hbox{\strut data}} & reference \\ 
\rule{0pt}{4pt}\\
\hline
\rule{0pt}{4pt}\\
89 & Ac & $6d7s^2$& 10.1& 193.3& 203.3 & 217 & calc. \cite{Miller}\\
 & & $7s^27p^1$ & 10.1& 131.8& 141.9& $-$ & $-$ \\
\rule{0pt}{2pt}\\
\hline
\rule{0pt}{2pt}\\
91 & Pa & $5f^26d7s^2$& 3.8& 150.6& 154.4& 171& calc. \cite{Miller}\\
 & & $5f^26d^27s^1$& 3.8& 148.1& 151.9& $-$ & $-$ \\
\rule{0pt}{2pt}\\
\hline
\rule{0pt}{2pt}\\
92 & U & $5f^37s^26d$& 3.8& 124.0& 127.8& 137(10)& exp. \cite{Kadar:1994}\\
 & & $5f^47s^2$& 4.3& 148.9& 153.2& 152.7 & calc. \cite{Miller} \\
\rule{0pt}{2pt}\\
\hline
\rule{0pt}{2pt}\\
93 & Np & $5f^46d7s^2$& 4.8 & 145.7 & 150.5 & 167 & calc. \cite{Miller}\\
 & & $5f^57s^2$& 5.8& 121.7& 127.5& $-$ & $-$ \\
\rule{0pt}{2pt}\\
\hline
\rule{0pt}{2pt}\\
94 & Pu & $5f^67s^2$ & 6.5 & 125.7 & 132.2 & 165 & calc. \cite{Miller}\\
 & & $5f^56d7s^2$& 5.2& 142.4& 147.6& $-$ & $-$\\
\rule{0pt}{2pt}\\
\hline
\rule{0pt}{2pt}\\
95 & Am & $5f^77s^2$& 7.2 & 124 & 131.2 & \vtop{\hbox{\strut 157}\hbox{\strut 116}} & \vtop{\hbox{\strut calc. \cite{Miller}}\hbox{\strut calc. \cite{Kello:1992}}}\\
 & & $5f^66d7s^2$& 5.4& 139.3& 144.7& $-$ & $-$\\
\rule{0pt}{2pt}\\
\hline
\rule{0pt}{2pt}\\
96 & Cm & $5f^76d7s^2$& 5.6& 137& 143.6 & 155 & calc. \cite{Miller}\\
 & & $5f^87s^2$& 7.6 & 121& 128.6& $-$ & $-$ \\
\rule{0pt}{2pt}\\
\hline
\rule{0pt}{2pt}\\
97 & Bk & $5f^97s^2$& 8& 117.3 & 125.3 & 153 & calc. \cite{Miller}\\
 & & $5f^86d7s^2$& 5.8& 135.8& 141.6& $-$ & $-$\\
\rule{0pt}{2pt}\\
\hline
\rule{0pt}{2pt}\\
98 & Cf &$5f^{10}7s^2$ & 8.2 & 113.3 & 121.5 & 138 & calc. \cite{Miller}\\
 & & $5f^{9}6d7s^2$& 5.8 & 136.5 & 142.3& $-$ & $-$\\
\rule{0pt}{2pt}\\
\hline
\rule{0pt}{2pt}\\
99 & Es &$5f^{11}7s^2$ & 8.3 & 109.2 & 117.5 & 133 & calc. \cite{Miller}\\
 & & $5f^{10}6d7s^2$& 5.9 & 140.2 & 146.1& $-$ & $-$\\
\rule{0pt}{2pt}\\
\hline
\rule{0pt}{2pt}\\
100 & Fm & $5f^{12}7s^2$& 8.4 &105 & 113.4 & 161 & calc. \cite{Miller}\\
 & & $5f^{11}6d7s^2$& 6 & 149.6 & 155.6& $-$ & $-$\\
\rule{0pt}{2pt}\\
\hline
\rule{0pt}{2pt}\\
101 & Md &$5f^{13}7s^2$& 8.5& 100.9 &109.4 & 123 & calc. \cite{Miller}\\
 & & $5f^{12}6d7s^2$& 6 & 173.6 & 179.6& $-$ & $-$\\
\rule{0pt}{2pt}\\
\hline
\rule{0pt}{2pt}\\
102 & No &$5f^{14}7s^2$& 8.5 & 96.9 & 105.4 & \vtop{\hbox{\strut 118}\hbox{\strut 110.8}} & \vtop{\hbox{\strut calc. \cite{Miller}}\hbox{\strut calc. \cite{Tierfelder}}}\\
 & & $5f^{14}7s^17p^1$& 8.5 & 259.3 & 267.8& $-$ & $-$\\
\rule{0pt}{4pt}\\
\hline\hline
\end{tabular}}\label{Tab:2}
\end{table*}

\section{Configuration interaction calculation of polarizabilities}
\label{Method}
\subsection{CI+MBPT calculation of polarizabilities}.

We have demonstrated in previous section that calculation of
polarizabilities of atom with open $f$-shell can be reduced to the
calculation of polarizabilities for atoms with two or three valence
electrons which form the $6s^2$ or $6s^25d$ configurations in
lanthanides and $7s^2$ or $7s^26d$ configurations in actinides. The
open $4f$ or $5f$ shell is attributed to the core and treated as it is
fully occupied but its contribution to the potential is rescaled with
the fractional occupation number.

The calculations are performed with the use of the CI+MBPT method.
Detailed description of the method  can be found in our earlier works
\cite{Dzuba:1996, Dzuba:2005, Ginges:2006}. A brief description of
this method is presented in this section.    

We use the $V^{N-M}$ approximation~\cite{Dzuba:2005}. The core
electron states are obtained in Hartree-Fock approximation for $N-M$
electrons, where $N$ 
and $M$ are total number of electrons and number of electrons above
closed shells ("valence electrons"), excluding the $f$-shell electrons.
Contribution of the latter is included in self consistent
potential of the core with "weight", fractional occupation number that
is equal to the ratio of $n/14$, where $n$ is number of $f$-shell
electrons. The Hartree-Fock (HF) Hamiltonian of 
the system has the form
\begin{equation}\label{HF}
\hat H_{HF}(r_i)=c\alpha \hat {\bf p_i} +
(\beta-1)mc^2-\frac{Ze^2}{r_i}+V^{N-M}(r_i), 
\end{equation}
where ${\bf \hat p_i}$ and ${\bf r_i}$ are operator of momentum and
coordinate of electron, $V^{N-M}$ is the self-consistent potential of the core.

Many-electron states for valence electrons can be obtained using the CI
and MBPT methods. The effective CI Hamiltonian has the form 
\begin{equation}\label{H_eff}
\hat H^{CI}=\sum_{i=1}^{M}\hat h_1(r_i)+\sum_{j>i=1}^{M}\hat h_2(r_i, r_j),
\end{equation}
where $\hat h_1(r)$ is the single-electron operator and $\hat h_2(r_i,
r_j)$ is the two-electron operator. The single electron operator $\hat
h_1(r)$ differs from (\ref{HF}) by an extra operator $\Sigma_1(r)$
\begin{equation}\label{h1}
\hat h_1(r_i)=\hat H_{HF}(r_i)+\Sigma_1(r_i). 
\end{equation}
This $\Sigma_1$ operator represents correlation interaction between a
particular valence electron and electrons in the core.
The two electron part of (\ref{H_eff}) is given by 
\begin{equation}\label{h2}
\hat h_2(r_i, r_j)=\frac{e^2}{|{\bf r_i}-{\bf r_j}|}+\Sigma_2(r_i, r_j),
\end{equation}
where $\Sigma_2$ accounts for screening of Coulomb interaction between
valence electrons by core electrons. For our purposes $\Sigma_1$ and
$\Sigma_2$ operators are sufficient to be accounted in the lowest,
second order of the MBPT.   

The CI many-electron wave function is written in a form
\begin{equation}\label{func}
\Psi=\sum_k c_k \Phi_k(r_1,...,r_M),
\end{equation}
where  $\Phi_k$ are determinants
made of single electron eigenfunctions of (\ref{HF}) combined in a way
to have appropriate value of total angular moment $J$. Here total
angular momentum $J$ is not the actual total angular momentum of the
$4f^{M-2}6s^2$ or $4f^{M-3}6s^25d$ configuration, but the total
angular momentum of smaller subsystem, e.g. $J=0$ for the $6s^2$
configuration of valence electrons and $J=3/2$ or $5/2$ for the $6s^25d$
configuration. The expansion coefficients $c_k$ and corresponding energies
are found by solving the matrix eigenvalue problem
\begin{equation} \label{CIM}
\hat  H^{CI}\Psi = E\Psi
\end{equation}
for lowest states of definite $J$ and parity.

Electric dipole transition amplitudes in (\ref{alpha0})
are calculated using the time-dependent Hartree-Fock
method~\cite{Dzuba:1987} (which is equivalent to the RPA method) and
the CI method   
\begin{equation}\label{E1}
\left<a|D_z|n\right>=\left<\Psi^{(a)}|d_z+\delta V^{N-M}|\Psi^{(n)}\right>,
\end{equation}
where $d_z=-ez$ is the $z$-component of the dipole moment operator and
$\delta V^{N-M}$ is the
correction to core potential due to its polarization by
external electric field. Electron wavefunctions $\Psi^{(a)}$ and
$\Psi^{(n)}$ were obtained using described above technique. 

To calculate scalar polarizabilities using formula (\ref{alpha0})
summation over complete set of
intermediate many-electron states needs to be carried out. We use the
Dalgarno-Lewis method~\cite{Dalgarno:1955} to 
reduce this summation to solving a system of linear equations with the
CI matrix. The expression for the polarizability (\ref{alpha0}) is rewritten as 
\begin{equation}
\alpha_0(a) = \frac{2}{3(2J_2+1)}\sum_{J_3=J_2\pm
  1}\left<\delta\Psi_{J_3}^{(a)}||{\bf 
    d}||\Psi_{J_2}^{(a)}\right>,\label{scl}
\end{equation}
where $J_2$ is total angular momentum of valence electrons.  
The correction $\delta\Psi_{J_3}^{(a)}$ to the wavefunction
$\Psi_{J}^{(a)}$ due to the laser electric field is found from the
matrix equation
\begin{equation}
\left(H^{CI}-E_a\right)\delta \Psi_{J_3}^{(a)} = -\left(d_z+\delta
  V^{N-M}\right)\Psi_{J_2}^{(a)}. 
\end{equation}


\subsection{CI calculations for systems with many valence electron.}

\label{CI-VN}
Previous consideration was based on the assumption that $f$-electrons
can be attributed to the core and the problem can be reduced to two or
three valence electrons above closed shells. This allows us to use
very advanced and accurate CI+MBPT method to perform the
calculations. It is useful however to check the calculations with an
alternative technique which is free from the assumption, even though
the technique is less accurate. In this section we move
$f$-electrons back to the valence space and use the CI technique which
treat them the same way as other valence electrons. The total number
of valence electrons for atoms with open $f$-shell varies between four
and sixteen. Below we consider examples of dysprosium, erbium and
thulium atoms in the $4f^{10}6s^2$, $4f^{12}6s^2$ and
$4f^{12}6s^25d$ configuration respectively. The number of valence
electrons is twelve for Dy, fourteen for Er and fifteen for Tm.
The use of the CI+MBPT method considered above is not possible for so
large number of valence electrons. We use an alternative CI technique
developed in our earlier works~\cite{Dzuba:2008, Flambaum:2008}.
This technique does not use excited single-electron states in the
basis. It tries instead to optimize the basis made of the lowest
single-electron states. The Hartree-Fock Hamiltonian used to construct
the basis has the form
\begin{equation}\label{HF-M}
\hat H_{B}=\sum_{i=1}^{N}c\alpha \hat {\bf p_i} +
(\beta-1)mc^2-\frac{Ze^2}{r_i}+V^{N}(r_i)
\end{equation} 
Here $V^N$ is the self-consistent Hartree-Fock potential created by
all atomic electrons. It is considered to be different for different
configurations of valence electrons (see Ref.~\cite{Dzuba:2008,
  Flambaum:2008} for details). The CI Hamiltonian has the form
\begin{eqnarray}\label{CI-M}
&&\hat H_{CI}=\sum_{i=1}^{M}\left[\vphantom{\sum_{i=1}^{M}}c\alpha
  \hat {\bf p_i} + 
(\beta-1)mc^2-\frac{Ze^2}{r_i}+V^{N-M}(r_i)+\nonumber\right.\\
&&\left.\delta V(r_i)\right] +\sum_{j>i=1}^{M}\frac{e^2}{|{\bf r_i}-{\bf r_j}|},
\end{eqnarray}
where $M$ is the number of electrons above closed shells. 
Apart from the number of valence electrons the CI Hamiltonian
(\ref{CI-M}) has two important differences from the CI+MBPT
Hamiltonian (\ref{H_eff}), the $\Sigma_2$ operator is not
included and the $\Sigma_1$ operator is approximated by a parametric
potential in a form
\begin{equation}\label{deltaV}
\delta V(r_i)=-\frac{\alpha_p}{2(r_i^4+a^4)}.
\end{equation}
Here $a$ is roughly core radius (we use $a=a_B$).
The form of (\ref{deltaV})
is chosen to match the long range polarization potential. 
Therefore, $\alpha_p$ is effectively core polarizability. It is
assumed to be different for different configurations. This allows us
to fit energy intervals between  states of different configurations by
treating $\alpha_p$ as a fitting parameter.  

Using this method to construct the wave function of the ground state
and hundreds of states for the summation in (\ref{alpha0}) and
using the RPA method to calculate amplitudes (\ref{E1}) we calculate
polarizabilities of many-electron atoms .This method was used for the
$4f^{10}6s^2$ configuration of Dy~\cite{Dy} and the $4f^{12}6s^2$
configuration of Er~\cite{Er}. In this work we reevaluated
polarizabilities of Dy and Er using extended basis and calculated the
polarizability of the first state of the $4f^{12}6s^25d$ configuration
of Tm ($E=13119.61 \ {\rm cm}^{-1}$). 
The results, $\alpha_0=165$ a.u. for Dy, $\alpha_0=169$ a.u. for Er,
and $\alpha_0=122$ a.u. for Tm are in good agreement with the results
reported in previous section. 


\section{Results}

The results for scalar polarizabilities of ground and first exited
configurations of lanthanides and actinides are presented in tables
\ref{Tab:1} and \ref{Tab:2}.
Total scalar polarizability is a sum of core electron (forth column)
and valence electron (fifth column) contributions. Core electron
contribution is calculated using the RPA approximation described in
section \ref{s:c-s}. Valence f-shell electrons contribution was
accounted in the core as closed f-shell with fractional occupation
number equal to $n/14$, where $n$ is the number of $f$-shell 
electrons. Contribution of remaining valence electrons presented in
fifth column in tables \ref{Tab:1} and \ref{Tab:2} were obtained using
the CI+MBPT method described in section~\ref{Method}. All presented 
values are in atomic units. In approximation used in this paper scalar
polarizabilities do not depend on the values of total angular momentum
as it was shown in section
\ref{f-elems}, therefore their values are the same for all levels of
a given configuration. Two last columns represent results from
\cite{Miller} and some other sources for comparison. As
one can notice, agreement is quite good although employed methods are
quite different. Extended estimate of accuracy together with 
comparison with available experimental measurements is presented in
next section.

\section{Discussion of accuracy}

\begin{table*}\center
\caption{Calculated and experimental static scalar polarizabilities of
  ytterbium (in a.u.).}
{\renewcommand{\arraystretch}{0}%
\begin{tabular}{l c l l l l l l}
\hline\hline
\rule{0pt}{4pt}\\
Z & element & State & core & \vtop{\hbox{\strut
    valence}\hbox{\strut (CI+MBPT)}} & \vtop{\hbox{\strut total
    scalar}\hbox{\strut polarizability}\hbox{\strut (core+valence)}} &
\vtop{\hbox{\strut existing}\hbox{\strut data}} & reference\\ 
\rule{0pt}{4pt}\\
\hline
\rule{0pt}{4pt}\\
70 & Yb & $4f^{14}6s^2 \ ^1$S$_0$& 6.4 & 132.5 & 138.9 & \vtop{\hbox{\strut
    142}\hbox{\strut 144.6}\hbox{\strut 140.7}\hbox{\strut
    141(6)}\hbox{\strut 142.6}\hbox{\strut $134.4(1.0)\le\alpha_0\le144.2(1.0)$}} &
\vtop{\hbox{\strut calc. \cite{Miller}}\hbox{\strut
    calc. \cite{Sahoo:2008}}\hbox{\strut
    calc. \cite{Tierfelder}}\hbox{\strut
    calc. \cite{Dzuba:2010}}\hbox{\strut
    calc. \cite{Buchachenko}}\hbox{\strut exp. \cite{Beloy:2010}}}\\ 
 & & $4f^{14}6s6p \ ^3$P$_0^o$& 6.4 & 305.8 & 312.2 & \vtop{\hbox{\strut
     315.9}\hbox{\strut 252(25)}\hbox{\strut 266(15)}\hbox{\strut
     302(14)}\hbox{\strut $280.1(1.0)\le\alpha_0\le289.9(1.0)$}} & \vtop{\hbox{\strut
     calc. \cite{Dzuba:2010}}\hbox{\strut
     calc. \cite{Porsev:1999}}\hbox{\strut
     calc. \cite{Porsev:2006}}\hbox{\strut
     calc. \cite{Dzuba:2010}}\hbox{\strut exp. \cite{Beloy:2010}}}\\ 
\rule{0pt}{2pt}\\
\hline\hline
\end{tabular}}\label{Tab:3}
\end{table*}

To estimate accuracy of present calculations we compare the results
obtained in different approaches used in this and earlier works.
We also compare the results with available experimental
data. There are strong indications that the accuracy of present
calculations is on the level of 15\% or better. 

Test calculations with the use of the many-valence-electrons CI method
described in section \ref{CI-VN} show no more than 13\% deviation from
the results presented in Table~\ref{Tab:1}. Given that that method is
likely to be less accurate than the main CI+MBPT method used in
present work, the actual accuracy of the results presented in Tables
\ref{Tab:1} and \ref{Tab:2} might be better.

We use expression (\ref{alpha2}) to calculate scalar
polarizabilities. Note that this expression does not depend on the
total angular momentum of the atom, $J_a$. It does not also depend on
the total angular momentum of the $f$-subshell, $J_1$. However, it
does depend on the total angular momentum of remaining valence
electrons, $J_2$, which is strictly speaking is not known. This does not
lead to a problem for the $4f^n6s^2$ configurations since $J_2=0$ for
the $6s^2$ configuration. If we consider the $4f^{n-1}6s^25d$
configuration instead, which is divided into the $4f^{n-1}$ and
$6s^25d$ subsystems, than there are two possibilities for the $6s^25d$
subsystem, $J_2=3/2$ and $J_2=5/2$. It is important to check that the
results are the same for both cases. We have done this test for
gadolinium atom. Calculations for the $4f^76s^25d$ configuration
assuming $J_2=3/2$ led to $\alpha_0=153.6$ a.u. (see
Table~\ref{Tab:1}) while calculations with $J_2=5/2$ gave
$\alpha_0=153.8$ a.u., the difference is about 0.1\%.
 
The most complete other theoretical data comes from the calculations
of Doolen~\cite{Miller}. Estimated accuracy of these calculations is
25\%. However, as one can see from Tables \ref{Tab:1} and \ref{Tab:2}
the agreement between two sets of results is significantly better for
most of atoms. It is about 10\% for lanthanides and slightly worse for
actinides. There is a special case of fermium atoms where the result
of Ref.~\cite{Miller} jumps to a high value breaking the trend along the
row of actinides. In contrast, the change in the value of scalar
polarizabilities for actinides is very smooth in our calculations. We
see no reason for fermium to be very different from its neighbors. The
difference between our results and those of Ref.~\cite{Miller} for
other actinides varies between 7 and 20\%, being smaller than 15\% for
most of atoms.

The most detailed study of the polarizabilities of lanthanides has
been done for ytterbium atom. This is because it has relatively simple
electron structure with fully filled $4f$ subshell and because it has
the $^1\rm{S}_0$ - $^3\rm{P}_0^o$ transition which is suitable for
atomic clocks. The available theoretical and experimental data for
these two states of ytterbium is summarized in Table~\ref{Tab:3}. Our
result for the ground state of Yb is within 5\% of other accurate
calculations and experimental limits found in
Ref.~\cite{Beloy:2010}. The result for the excited $6s6p \
^3\rm{P}^o_0$ state is less accurate but still within 12\% of other
accurate calculations and experimental limits.

Experimental data on static scalar polarizabilities of lanthanides and
actinides is absent. There are measurements of the dynamic
polarizabilities for dysprosium~\cite{Lev}, erbium~\cite{Ferlaino},
and uranium~\cite{Kadar:1994}. Scalar polarizability of uranium
interpolated to $\omega=0$ is 137(10) a.u.~\cite{Kadar:1994} which
differ by about 10\% from our calculated value of 153 a.u. (see
Table~\ref{Tab:3}). The situation for dysprosium and erbium is
different. Measured dynamic polarizabilities of both atoms are
significantly smaller than the calculated static polarizabilities. For
example, $\alpha_0(\lambda=1064 \ \rm{nm}) = 116$ a.u. for Dy~\cite{Lev},
and  $\alpha_0(\lambda=1064 \ \rm{nm}) = 84(2)(18)$ a.u. for
Er~\cite{Ferlaino}, while calculated static polarizabilities are 163
a.u. for Dy and 150 a.u. for Er (see Table~\ref{Tab:1}). If all
numbers are correct than the most likely explanation for the shift in
the polarizabilities is the presence of a strong resonance between
$\omega=0$ and $\omega=8398$ cm$^{-1}$ ($\lambda = 1064$ nm). There is
indeed resonances in both atoms which correspond to the $4f$ - $5d$
single-electron transitions. However, according to our estimations, the
amplitudes of the transitions between ground and resonance states are
too small to explain the difference between theory and
experiment. Another possible explanation relies on tensor
polarizability. If tensor polarizability is large, then depending on
the geometry of the measurements, the effective polarizability might
be small. However, here again our estimations show that tensor
polarizabilities of both atoms are too small to explain the
difference. In the end the reason for disagreement is not
clear. However, based on the arguments presented above, we belive that
the accuracy of our result is about 15\% or better.

\begin{acknowledgments}

The work is partly supported by the Australian Research Council.

\end{acknowledgments}

\end{document}